\documentclass[aps,twocolumn,prb,showpacs,multicol,amsmath,amssymb]{revtex4-1}
\usepackage[dvips]{graphicx}
\usepackage{braket}
\usepackage{color}
\newcommand{\be}{\begin{equation}}
\newcommand{\ee}{\end{equation}}

\newcommand{\bea}{\begin{eqnarray}}
\newcommand{\eea}{\end{eqnarray}}
\newcommand{\bd}{\begin{displaymath}}
\newcommand{\ed}{\end{displaymath}}
\newcommand{\ba}{\begin{array}}
\newcommand{\ea}{\end{array}}
\newcommand{\bi}{\begin{itemize}}
\newcommand{\ei}{\end{itemize}}
\newcommand{\bc}{\begin{center}}
\newcommand{\ec}{\end{center}}
\newcommand{\bfl}{\begin{flushleft}}
\newcommand{\efl}{\end{flushleft}}
\newcommand{\bfr}{\begin{flushright}}
\newcommand{\efr}{\end{flushright}}

\def\6{\partial}

\def\={\!\!\!&=&\!\!\!}
\def\+{\!\!\!&&\!\!\!+~}
\def\-{\!\!\!&&\!\!\!-~}

\begin{document}

\title[]{Polar Kerr effect from a time-reversal symmetry breaking \\unidirectional charge density wave}
\author {M. Gradhand$^{1}$, I. Eremin$^2$ and J. Knolle$^{3}$ }
 \affiliation{
 $^{1}$H. H. Wills Physics Laboratory, University of Bristol, Bristol BS8 1TL, United Kingdom}
 \affiliation{
 $^{2}$Institut f\"ur Theoretische Physik III, Ruhr-Universit\"at Bochum, Universit\"atsstrasse 150, DE-44801 Bochum, Germany}
 \affiliation{$^3$T.C.M. Group, Cavendish Laboratory, J.~J.~Thomson Avenue, Cambridge CB3 0HE, United Kingdom}

\begin{abstract}
We analyze the Hall conductivity $\sigma_{xy}(\omega)$ of a charge ordered state with momentum $\mathbf{Q}=(0,2Q)$ and calculate the intrinsic contribution to the Kerr angle $\Theta_K$ using the fully reconstructed tight-binding band structure for layered cuprates beyond the low energy hot spots model and particle hole symmetry. We show that such a unidirectional charge density wave (CDW), which breaks time reversal symmetry as recently put forward by Wang and Chubukov [Phys. Rev. B {\bf 90}, 035149 (2014)], leads to a nonzero polar Kerr effect as observed experimentally. In addition, we model a fluctuating CDW via a large quasiparticle damping of the order of the CDW gap and discuss possible implications for the pseudogap phase. We can qualitatively reproduce previous measurements of underdoped cuprates but making quantitative connections to experiments is hampered by the sensitivity of the polar Kerr effect with respect to the complex refractive index $n(\omega)$.
\end{abstract}

\date{\today}


\maketitle

%\section{Introduction}%%%%%%%%%%%%%%%%%%%%%%%%%%%
{\it Introduction.}
One of the most controversial topics in the field of high-temperature superconductivity is the origin of the so-called
“pseudogap” phenomenon observed by various experimental
techniques in the underdoped cuprates \cite{Alloul1989,Timusk1999,Kaminski2014} at temperatures $T^*$ being larger than the superconducting transition temperature, T$_c$. A large number of theoretical scenarios
have been initially proposed to explain the origin of the
pseudogap. \cite{theory_early}
Recent experimental studies of hole-doped cuprates, however, have indicated that the pseudogap region of the high-T$_c$ cuprates is a state, which competes with superconductivity, and also breaks several symmetries. \cite{huecker,tranquada,hinkov,hinkov2008,fujita,Xia_2008,He_2011,kapitulnik} In particular, x-ray, and neutron scattering experiments
as well as Scanning Tunneling Microscopy (STM) have found the breaking of the lattice symmetry from C$_4$ down to C$_2$ below the pseudogap temperature, T$^*$, in several cuprate compounds\cite{huecker,tranquada,hinkov,fujita}. In addition,  recent findings of the polar Kerr effect\cite{Xia_2008,He_2011,kapitulnik}, and
the intra-unit cell magnetic order\cite{bourges} point towards breaking of time-reversal or mirror symmetries in the underdoped cuprates.
Furthermore, direct indications in favour of the static incommensurate charge-density-wave (CDW)
order with momenta ${\bf Q}_x = (2Q,0)$ or ${\bf Q}_y = (0,2Q)$ were found by X-ray measurements\cite{ghiringelli,achkar,comin,silvaneto}, nuclear magnetic resonance technique\cite{taowu1,taowu2} and in experiments on the sound velocity in a magnetic field\cite{leboeuf}.  In these systems $2Q$ refers to the distance between the so-called hot spots on the Fermi surface, i.e. points where the magnetic Brillouine Zones intersects the Fermi surface (FS). It was also argued that the charge
order has predominantly $d$-wave form
factor\cite{fujita} and that the formation of Fermi surface arcs coincides with the formation of CDW order\cite{Fuji2014}.

Possible explanations of the charge order and related other symmetry breaking include loop-current order\cite{varma} or $d$-density-wave
(current) order\cite{chakravarty}. Within the so-called spin fluctuation scenario, which supports also $d-$wave superconductivity, there is another instability in the $d$-wave charge channel, associated with the wavevector ${\bf Q}_d = 2{\bf k}_{hs}$\cite{sachdev}. Here, ${\bf k}_{hs}$ is the momentum of one of the hot spots such that $ 2{\bf k}_{hs} = (±2Q, ± 2Q)$ points along one of the Brillouin Zone diagonals. Efetov, Meier, and Pepin\cite{Efetov2013,Meier2014} argued that the pseudogap can be the consequence of the competition between $d$-wave charge order (bond order) and
superconductivity (SC). In this case $T^*$ refers to the formation of the combined SC/CDW “supervector” order parameter, but its direction gets fixed along the SC “axis” only at lower temperature  T$_{C}$. Sachdev and collaborators\cite{Sachdev2013,Allais2014,Sau2014} proposed that magnetically-mediated interaction can also yield an attraction in the CDW channel for unidirectional incoming momenta $\mathbf{Q}_x=(0,2Q)$ or $\mathbf{Q}_y=(2Q,0)$, which has also been found in three band models~\cite{Bulut2013,Sachdev2014b}.
In these proposals the ground states break $C_4$ rotational symmetry of the lattice. Most recently Wang and Chubukov~\cite{Chubukov2014} have shown that for large magnetic correlation length there could be an additional CDW instability with both real and imaginary order parameter at a nonzero temperature breaking also time reversal symmetry. Translated into real space the latter order has both bond and site charge density and bond current modulations.

\begin{figure}[]
\begin{centering}
\includegraphics[width=1.0\columnwidth]{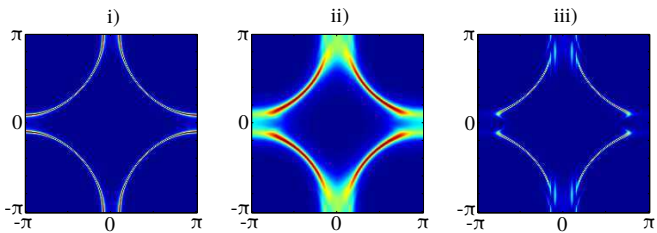}
\par\end{centering}
\caption{The spectral function $I(\mathbf{k},\omega=0)$ is shown in the full Brillouin zone for three different cases (from left to right): i) normal state, $\Delta_1=\Delta_2=0$ and small quasiparticle damping $\Gamma=10$~meV; ii) fluctuating CDW state, $\Delta_1=\Delta_2=50$~meV modelled by a large damping $\Gamma=50$~meV; iii) long range ordered CDW state, $\Delta_1=\Delta_2=50$~meV ($\Gamma=10$~meV). \label{QPI_ARPES_n200}}
\end{figure}
\begin{figure}
\begin{centering}
\includegraphics[width=0.90\linewidth]{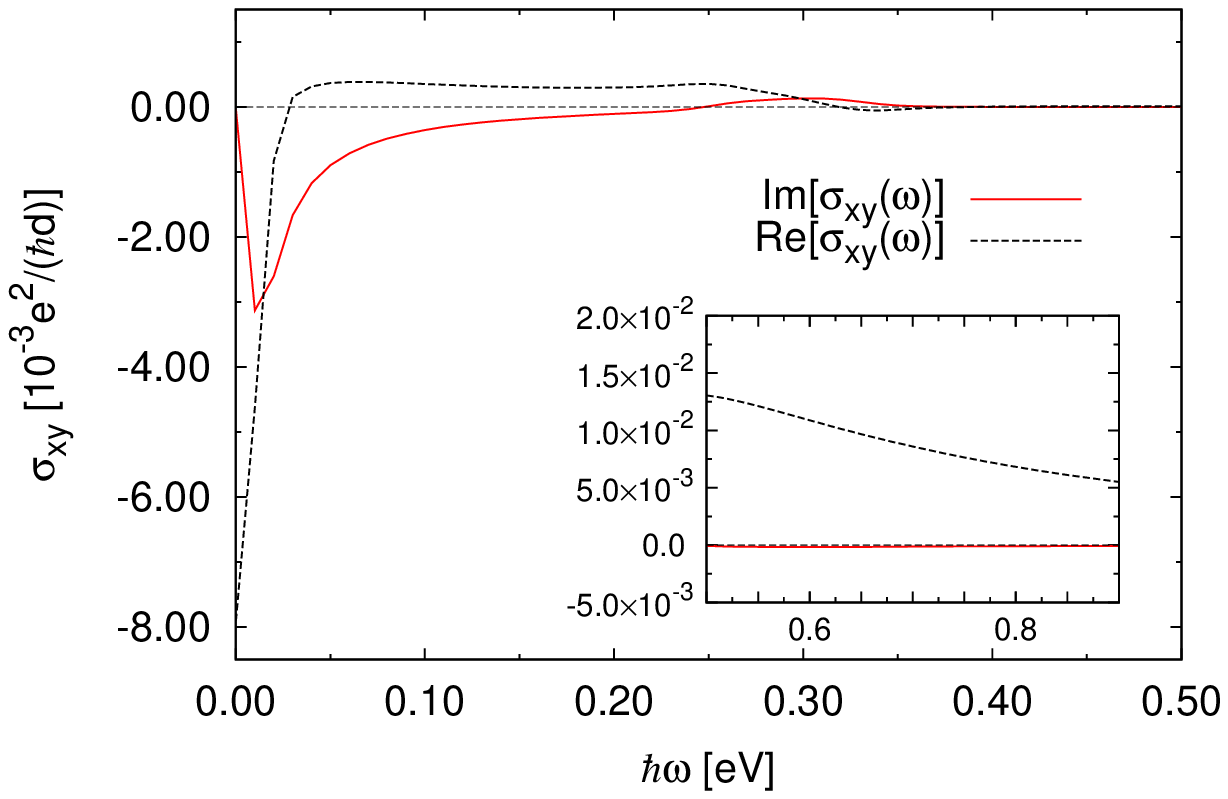}
\includegraphics[width=0.9\linewidth]{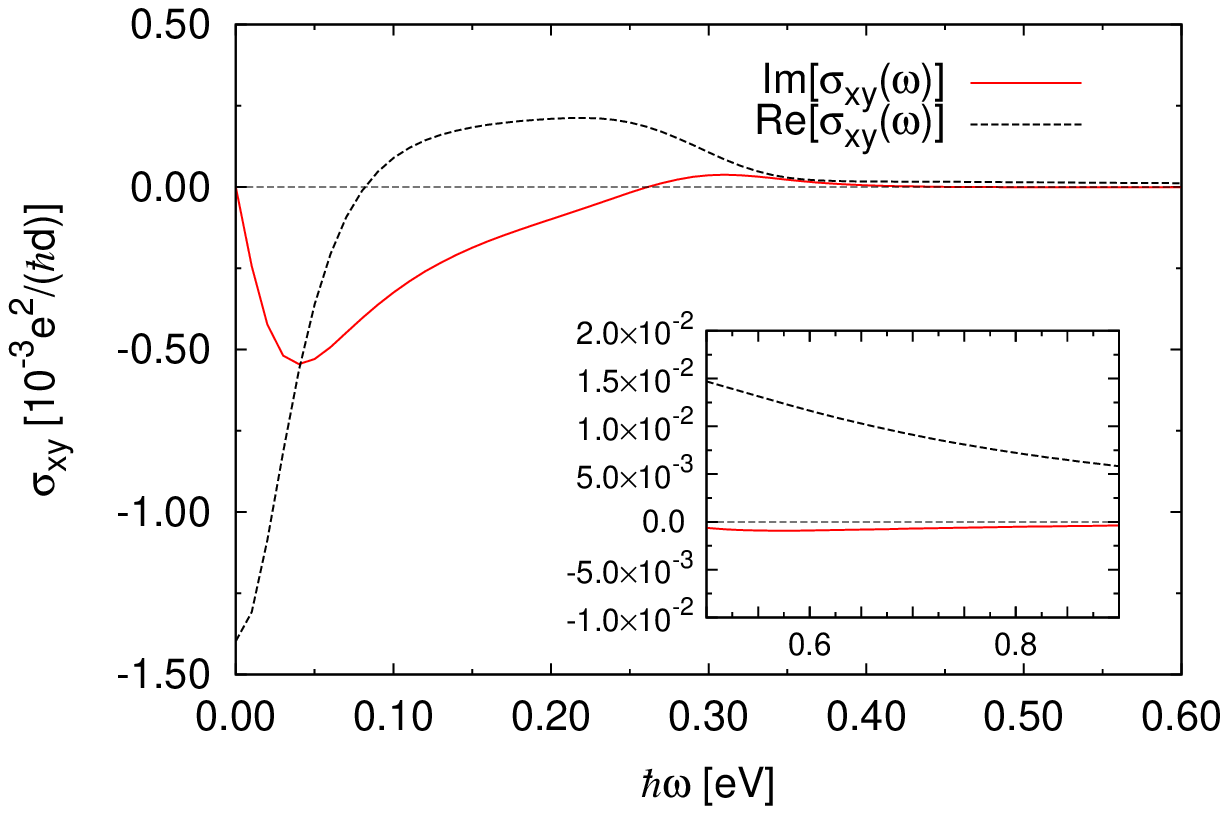}
\par\end{centering}
\caption{The real and imaginary part of the optical Hall conductivity as a function of frequency. The impurity broadening is $\Gamma=10$~meV and $\Gamma=50$~meV in the upper and lower panel, respectively. The insets show the relevant frequencies ($0.8$~eV) at which experiments were performed so far \cite{Xia_2006,Xia_2008,He_2011,kapitulnik}.     \label{Hall_cond}}
\end{figure}
In this Communication, we follow Ref.~\onlinecite{Chubukov2014} and analyze the Hall conductivity which arises from such a unidirectional time-reversal symmetry breaking CDW phase.
We do not attempt to answer the question whether or not this state truly minimizes the free energy of a particular model but concentrate on the experimental consequences once the state is formed. 
Our main objective is to study the intrinsic contributions  of the full lattice band structure to the polar Kerr effect, which is well established experimentally in the pseudogap phase~\cite{Xia_2008,He_2011,kapitulnik}. Its microscopic origin is still poorly understood and under current debate. Two original proposals \cite{Pershoguba_2013,Hosur_2013} to explain the nonzero Kerr signal based on inversion symmetry breaking were retracted \cite{Pershoguba_2014,Hosur_2014} and in a number of articles the necessity of time reversal symmetry breaking was emphasized.\cite{Orenstein_2013,Pershoguba_2014,Hosur_2014,Wang_2014} Note that our present study is complementary to a very recent work\cite{Wang_2014} which, however, analyzed only the extrinsic contributions to the Polar Kerr effect in a similar setting but restricted to an effective low energy description keeping only states around the hot spots.

{\it Theory.}
Our starting point is the CDW mean-field Hamiltonian
\begin{eqnarray}
\label{Hamiltonian}
 \hat{H} & = &  \sum_{\mathbf{k}, \sigma} \left\lbrace \varepsilon_{\mathbf{k}} c_{\mathbf{k} \sigma}^{\dagger} c_{\mathbf{k} \sigma} + \Delta^{\mathbf{Q}}_{\mathbf{k}} c^{\dagger}_{\mathbf{k} \sigma} c_{\mathbf{k}+2\mathbf{Q}} +\text{h.c.}\right\rbrace
 \end{eqnarray}
with  $\Delta^{\mathbf{Q}}_{\mathbf{k}}  =  \Delta_1 \cos{\left( k_x-k_0\right) } \cos{Q_y}-i \Delta_2 \sin{\left( k_x-k_0\right) }\sin{Q_y}$
 and  $2Q_y=\frac{\pi}{5}$ and  $k_0=0.9\pi$.
We use the parameters of the tight-binding energy dispersion from Ref.\cite{He_2011} to obtain a commensurate CDW. Note that the above order parameter satisfies the following symmetry relations $\left( \Delta^{\mathbf{Q}}_{\mathbf{k}}\right)^*=\Delta^{\mathbf{-Q}}_{\mathbf{k}}$ and transforms as $\left( \Delta^{\mathbf{-Q}}_{\mathbf{-k}}\right)^*=\Delta^{\mathbf{Q}}_{\mathbf{-k}}$ under time reversal symmetry. The centre of mass momentum of the CDW is at $k_0$ such that $\mathbf{k_{\text{hs}}}=(\pm k_0,\pm Q_y)$. This stripe-type order breaks both the rotational and the time-reversal symmetries. Since $\Delta^{\mathbf{Q}}_{\mathbf{k+2Q}}\neq \Delta^{\mathbf{Q}}_{\mathbf{k}}$ it is necessary to diagonalize a $10 \times 10$ matrix with all multiples $N 2 Q_y$ up to $N=10$ that closes the periodic Brillouin zone (BZ).
Since the CDW order parameter has two components there is the interesting possibility\cite{Chubukov2014} that a composite Ising like order breaking only  rotational and time reversal symmetry forms above $T_{CDW}$ at which long range translational order (LRO) gets broken. In the following we do not explicitly calculate such a fluctuating CDW but we mimick the effect by a large quasiparticle damping, $\Gamma$,  of the order of the CDW gap itself, $\Delta\approx \Gamma$.

We first investigate the direct feedback of charge order on the electronic structure which have been shown to be intimately connected\cite{Fuji2014}.
Angle resolved photo emmision spectroscopy (ARPES) measures directly the electronic spectral function $I(\mathbf{k},\omega) = \text{Im} \left[ \hat G^0(\mathbf{k},\omega)\right]_{1,1}$ which can be calculated from the bare Greens function:
\begin{eqnarray}
\label{GFbare}
\hat G^0 (\mathbf{k},\omega) & = & \left( \omega +i\Gamma - \hat H(\mathbf{k})\right)^{-1} .
\end{eqnarray}
We have omitted the spin label, $\hat H(\mathbf{k})$ is the $10 \times 10$ Hamiltonian matrix and $\Gamma$ is our phenomenological quasiparticle damping.
The results are shown in Fig.~\ref{QPI_ARPES_n200} for three different sets of parameters: (a) normal state FS, (b) the fluctuating CDW state modeled by $\Delta_1=\Delta_2=50 $~meV and a large damping $\Gamma=50$~meV, (c) the LRO CDW state with the same set of $\Delta_i$ but with a smaller damping $\Gamma=10$~meV.

The formation of the CDW gaps out states at the hot spots $\mathbf{k}_{hs}$ which together with the suppressed spectral weight outside the AFM BZ leads to the observed Fermi arc structure of the spectral function.
In accordance with ARPES experiments\cite{comin} the wave vector connecting the tips of the arc is larger than the wave vector connecting the hot spots of the normal state FS. The effect appears naturally in our mean-field description in which a full region around the hot spots is gapped out by the CDW order. However,  we have chosen the momentum of our CDW to match the vector connecting two hot spots\cite{Chubukov2014}, which could likely turn out to be different when investigating the origin of the ordered state beyond weak coupling. As this question is beyond the scope of our study we concentrate on the experimental consequences of the formed CDW phase. Our following results do not depend sensitively on the precise value of the ordering vector.

Next we study the polar Kerr effect which is quantified by the Kerr angle $\theta_K$. It can be expressed in terms of the Hall conductivity $\sigma_{xy}(\omega)$ and the complex refractive index $n(\omega)$ as \cite{Lutchyn_2009}
\begin{equation}\label{Eq:Kerr}
\theta_K=\frac{1}{\epsilon_0\omega}\text{ Im}\left[\frac{\sigma_{xy}(\omega)}{n(\omega)(n^2(\omega)-1)}\right]\ \text{,}
\end{equation}
where for a complex refractive index both the real and imaginary part of the Hall conductivity contribute. Generally, the Hall conductivity has an extrinsic and intrinsic contribution \cite{Onoda_2006,Sinitsyn_2008,Xiao_2010, Nagaosa_2010,Wang_2014}, where the latter can be expressed entirely in terms of the electronic structure of the unperturbed crystal. \cite{Karplus_1954} In contrast, the extrinsic contribution is mediated via impurity scattering. \cite{Smit_1958, Sinitsyn_2008, Kim_2008, Goryo_2008, Wang_2014} Fundamentally, the system needs to break mirror and time reversal symmetry to exhibit a finite intrinsic or extrinsic Hall conductivity.\cite{Kleiner_1969, Wang_2014} Furthermore, in order to create an intrinsic contribution multiple bands have to be considered. In a single band system the effect would vanish by definition. \cite{Taylor_2012, Gradhand2013} In the present case of a charge density wave, which breaks time reversal and $C_4$ rotational symmetry, the gap parameter is providing all conditions intrinsically and creates an effective multi-band system in the reduced Brillouin zone.

In general, the antisymmetric part of the Hall conductivity can be expressed as\cite{Gradhand2013}
\begin{widetext}
\begin{eqnarray}
\sigma_{xy}(\omega)&=&\frac{e^2 \hbar}{Vm^2}\sum\limits_{n,n^\prime, {\bf k}}f(E_{n}({\bf k})\left[1-f(E_{n^\prime}({\bf k}))\right]\times \nonumber \\
&\times & \frac{\text{Im}\left[u^{\dagger}_ {n^\prime}({\bf k}) H^x_{\bf k} u_{n}({\bf k})u^{\dagger}_{n}({\bf k})H^y_{\bf k}u_{n^\prime}({\bf k})-u_{n^\prime}({\bf k})H^y_{\bf k}u_{n}({\bf k})u^\dagger_{n}({\bf k}) H^x_{\bf k}u_{n^\prime}({\bf k})\right]}{(E_{n^\prime}({\bf k})-E_{n}({\bf k}))^2-(\hbar\omega+i\Gamma)^2}\ \text{,}\label{Eq:Re_cond}
\end{eqnarray}
\end{widetext}
where $u^\dagger_{n}({\bf k})=(u^1_{n}({\bf k}),u^2_{n}({\bf k}),...,u^{10}_{n}({\bf k}))$ are the eigenfunctions of the ${\bf k}$ dependent Hamilton matrix $\hat{H}({\bf k})$. The operator $H^{x,y}_{\bf k}$ is the ${\bf k}$ derivative with respect to $k_x$ or $k_y$, respectively. For the tight-binding model under consideration the important matrix elements can be written as
\begin{widetext}
\begin{equation}\label{Eq:Hint_TB}
\braket{n^\prime{\bf k} |H^{x,y}_{\bf k}|n{\bf k}}=\left(
\begin{array}{c}
u^1_{n^\prime}({\bf k})\\
u^2_{n^\prime}({\bf k})\\
\vdots
\end{array}
\right)^\dagger
\left(
\begin{array}{ccc}
\nabla_{{\bf k}_{x,y}}\epsilon_{\bf k}& \nabla_{{\bf k}_{x,y}}\Delta_{\bf k}^{\bf Q}&\hdots\\
\nabla_{{\bf k}_{x,y}}(\Delta_{\bf k}^{\bf Q})^\ast&\nabla_{{\bf k}_{x,y}}\epsilon_{{\bf k}-2{\bf Q}}&\hdots\\
\vdots&\vdots&\ddots
\end{array}
\right)\left(
\begin{array}{c}
u^1_{n}({\bf k})\\
u^2_{n}({\bf k})\\
\vdots
\end{array}
\right)\ \text{,}
\end{equation}
\end{widetext}
where the main assumption is that all matrix elements of the position operator vanish. \cite{Gradhand2013} With that at hand the Hall conductivity can be evaluated and the results are shown in Fig.~\ref{Hall_cond} for different broadening accounting for random impurity scattering.

For the considered model Hamiltonian, the antisymmetric part of the Hall conductivity will vanish in case of a real order parameter not breaking time reversal symmetry. Hence, the breaking of $C_4$ symmetry is not sufficient to induce a Kerr signal. The reason is straightforward and follows from arguments used for the existence or non-existence of the Berry curvature of a generic parameter-dependent Hamiltonian. In case of a real Hamiltonian the wave functions can be chosen to be real, leading to a vanishing Berry curvature for all {\bf k} points, except for degeneracies in the electronic bands.\cite{Berry_1984,Bohm_2003} Since the real part of the optical conductivity at zero frequency resembles the Berry curvature and is connected via Kramers Kronig transformation to the imaginary part both have to vanish~\cite{Gradhand_2014, Souza_2008}. The argument can be extended to finite frequencies as well. Only for a complex order parameter the considered model will exhibit a finite Berry curvature i.e. a finite optical Hall conductivity. In the following we will discuss the numerical results for the time reversal symmetry breaking order parameter in detail.
\begin{figure}
\begin{centering}
\includegraphics[width=0.9\linewidth]{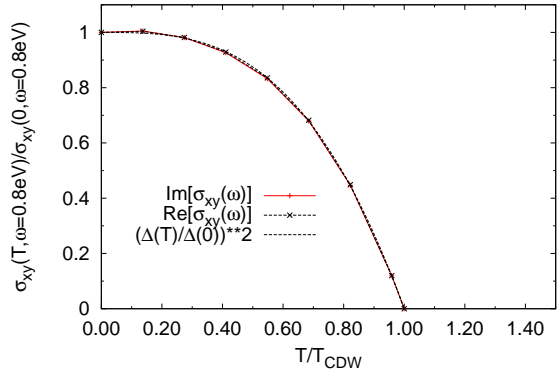}
\par\end{centering}
\caption{The normalized real and imaginary part of the optical Hall conductivity as a function of temperature ($\Gamma=10meV$). The temperature dependence of the gap is modeled mean-field-like using the function $\Delta(T)=\sqrt{1-(T/T_{CDW})^3}$.  \label{Hall_cond_Temp}}
\end{figure}

{\it Results.}
Fig.~\ref{Hall_cond} summarizes the the real and imaginary part of the antisymmetric Hall conductivity (Eq.~(\ref{Eq:Re_cond})) for the considered gap function and two distinct impurity scattering amplitudes. The signal is already nonzero at a frequency of the order parameter $\Delta_1=\Delta_2=5$~meV which is distinct from the results for Sr$_2$RuO$_4$. \cite{Taylor_2012, Gradhand2013} For the latter, the signal appeared at the interorbital coupling strength. Nevertheless, both results are not in contradiction to each other but in the present case the finding is merely an artefact of the fact that the multi-band structure and the symmetry breaking are both induced by the gap function. Furthermore, the signal is two orders of magnitude larger than for the realistic band structure of Sr$_2$RuO$_4$. \cite{Gradhand2013} It results from the order parameter being  one order of magnitude larger since it enters the conductivity quadratically. \cite{Taylor_2012} In the inset of Fig.~\ref{Hall_cond} we present the frequency region where the Kerr effect experiments on Sr$_2$RuO$_4$ and the cuprates   were performed.~\cite{Xia_2006,Xia_2008,He_2011,kapitulnik} Here, the signal drops by three orders of magnitude and the imaginary part almost vanishes. In contrast to Sr$_2$RuO$_4$, in the cuprates it might be feasible to perform corresponding experiments at lower frequencies due to the significantly higher energy scales (e.g. critical temperatures) and the much larger predicted signal. Extending the measurements to a larger frequency range would allow for more quantitative comparison between theory and experiments.

\begin{figure}
\begin{centering}
\includegraphics[width=0.9\linewidth]{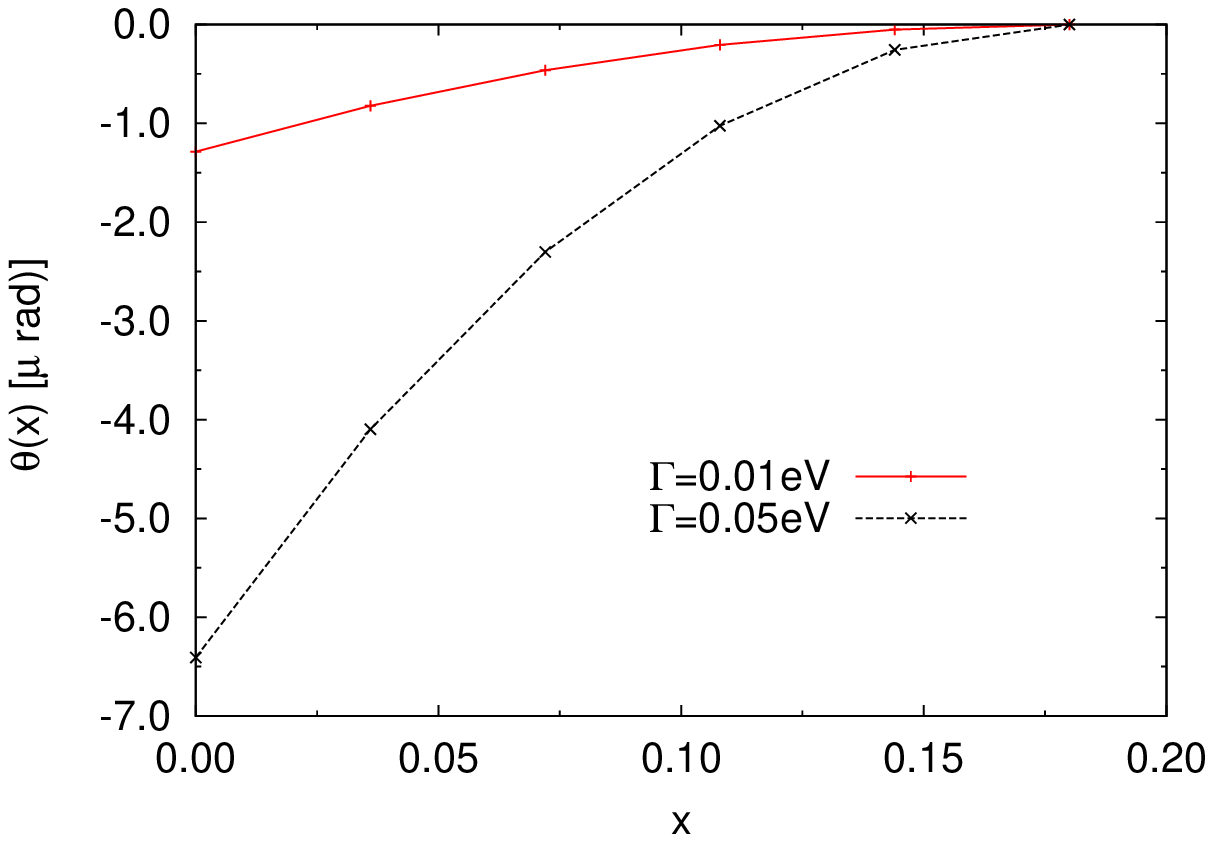}
\includegraphics[width=0.9\linewidth]{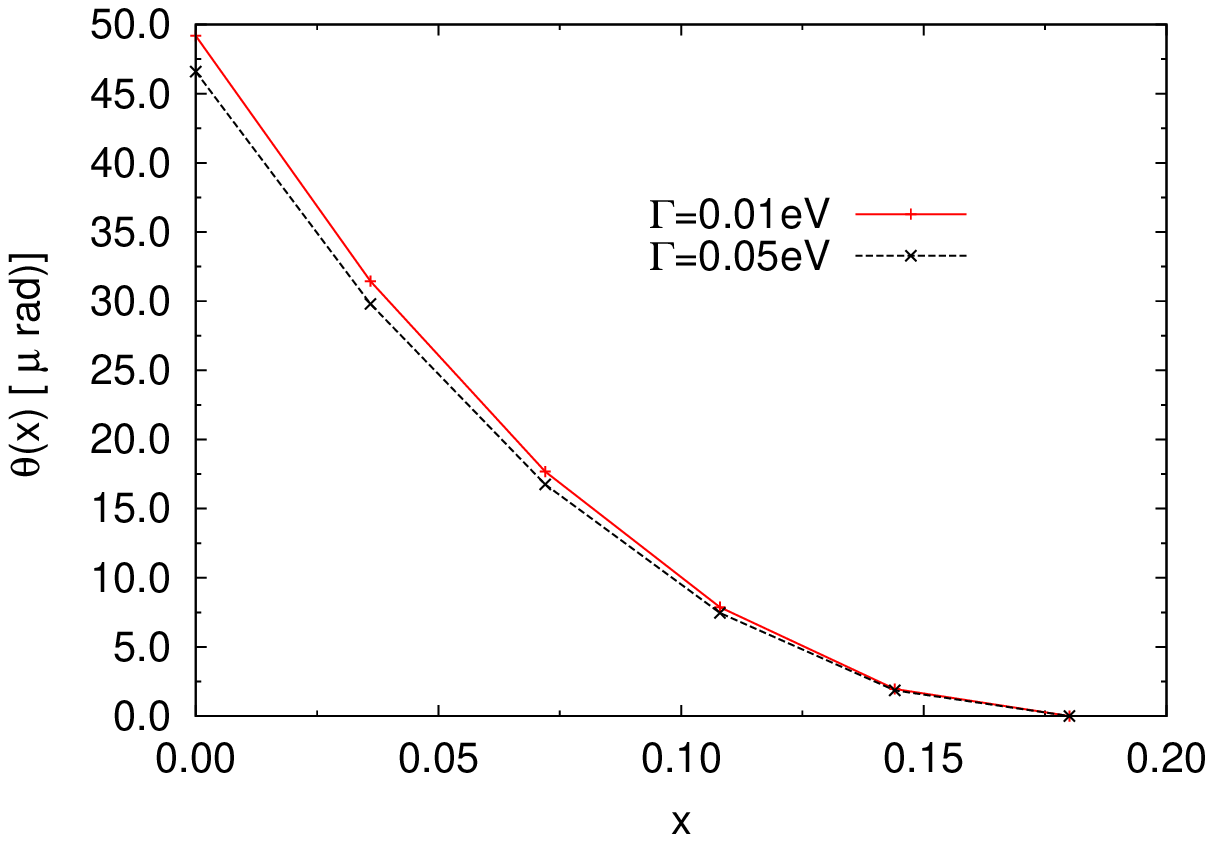}
\par\end{centering}
\caption{The Kerr angle as a function of doping $x$. The doping is modelled with a linear dependence of the gap on the doping as $\Delta(x)=\Delta(x=0)*(1-x/0.18)$~\cite{Tewari2008}. For the upper panel we assumed the refractive index to be pure real (n=1.692) in accordance to Ref.~\cite{Tewari2008}. In the lower panel we assumed a complex refractive index (taken from Ref.~\cite{Kezuka1991}) as   (n=1.692-i*0.403), leading to a dominance of the real part of the optical Hall conductivity. Note the change of sign between the two different assumptions. \label{Kerr_angle}}
\end{figure}
To model the temperature dependence of the optical conductivity we perform additional calculations assuming that the gap function follows a mean-field like behaviour as $\Delta(T)=\sqrt{1-(T/T_{CDW})^3}$. The result is shown in Fig.~(\ref{Hall_cond_Temp}), for the optical frequency of $0.8\ eV$. The temperature dependence of the conductivity follows the square of the order parameter as expected in such a simple model case. \cite{Taylor_2012}

Finally, we performed doping dependent calculations for the optical Kerr effect exploiting the results for Eq.~(\ref{Eq:Re_cond}) in combination with Eq.~(\ref{Eq:Kerr}). We follow earlier work~\cite{Tewari2008} and model the doping dependence with a phenomenological function $\Delta_{1/2}(x)=\Delta_{1/2}(x=0)*(1-x/0.18)$. For the refractive index we consider two separate approximations. In the first case (upper panel of Fig.~{\ref{Kerr_angle}) we assumed it to be purely real as n=1.692 according to Ref.~\cite{Tewari2008}. In that case the imaginary part of the optical conductivity is dominating the signal, which is not only small but strongly depends on the considered impurity broadening. On the other hand, assuming a more general complex refractive index\cite{Kezuka1991} as $n=1.692-i*0.403$  (lower panel of Fig.~\ref{Kerr_angle}) leads to the dominance of the real part of the optical conductivity in the Kerr angle. The significantly larger Kerr angle is furthermore almost independent of the impurity broadening. Nevertheless, the observed sign change of the Kerr angle between the two approximations highlights the importance of an accurate knowledge of the frequency dependent complex refractive index for qualitative but especially quantitative predictions.
Furthermore, we would like to point out that in general the Hall conductivity is a sum of intrinsic and extrinsic contributions. Our calculations rely on the intrinsic effect in contrast to Ref.~\onlinecite{Wang_2014} which focused on the extrinsic ones. Although, any conclusion about the dominance of either one contribution is ambiguous (especially at elevated frequencies) the intrinsic effect will be dominant for rather impure samples because of the impurity scaling behavior of the extrinsic mechanism (skew scattering).\cite{Onoda_2006,Lowitzer_2010}

Despite the mentioned obstacles preventing a determination of quantitatively reliable results our calculated Hall angle is several orders of magnitude larger than in the superconducting state of Sr$_2$RuO$_4$ which is in qualitative agreement to the experimental findings of Ref.~\onlinecite{Xia_2008,He_2011,kapitulnik}. Most strikingly, our results in the ordered phase ($\Gamma=10$~meV) with a purely real refractive index nicely reproduce the correct order of magnitude of the polar Kerr angle in all three cuprates YBa$_2$Cu$_3$O$_{6+x}$\cite{Xia_2008}, Pb$_{0.55}$Bi$_{1.5}$Sr$_{1.6}$La$_{0.4} $CuO$_{6+\delta}$~\cite{He_2011}, and  La$_{1.875}$Ba$_{0.125}$CuO$_4$\cite{kapitulnik} which are all of the order of $\theta_K\approx 1\mu$rad.

{\it Conclusion.}
In summary, we have shown that the unidirectional CDW order parameter recently proposed by Wang and Chubukov\cite{Chubukov2014} to account for time reversal symmetry breaking in the pseudogap phase, is capable to explain the polar Kerr effect measurements semi quantitatively. We have explicitly evaluated the antisymmetric part of the conductivity tensor for a model band structure describing the generic Fermi surface of the underdoped cuprates which allows us to numerically calculate the full frequency range of the optical conductivity.
Our results show that for a reliable quantitative comparison of experimental data and different theoretical scenarios a precise knowledge of the (in general complex) refractive index is called for. Together with further measurements at lower frequencies, most desirable at the order of the spectral gap itself, it has the potential to turn Polar Kerr effect measurements into a more powerful tool for elucidating the complicated many-body state of underdoped cuprates.

We thank A. Chubukov, K.B. Efetov and P.A. Volkov for helpful discussions. M.G. acknowledges financial support from the Leverhulme Trust via an Early Career Research Fellowship (ECF-2013-538). The work of J.K. is supported by a fellowship within the Postdoc-Program of the German Academic Exchange Service (DAAD).

\end{document}